 \documentclass[final,3p,times]{elsarticle}

\usepackage{amssymb}
\usepackage{mathrsfs,amssymb,amsthm,stmaryrd,amsmath,latexsym,indentfirst,color}
\usepackage{stmaryrd}
\usepackage{algorithm}
\usepackage{graphicx}
\usepackage{amsmath}
\usepackage{tabularx}
\usepackage{geometry}
\usepackage{booktabs}
\usepackage{epsfig}
\usepackage{epstopdf}
\usepackage{algorithmic}
\usepackage{colortbl}
\usepackage{multirow}  
\usepackage{enumitem}
\journal{\bf }
\usepackage{subfig}
\usepackage{romanbar}
\usepackage{rotating}
\usepackage{pifont} 

\begin{document}
\begin{frontmatter}

\title{{\bf\LARGE Privacy-preserving formal concept analysis: A homomorphic encryption-based concept construction}}

\author[km1,km2]{Qiangqiang Chen}

\author[km1,km2]{Yunfeng Ke}

\author[km3]{Shen Li\corref{cor1}}

\author[km1,km2]{Jinhai Li\corref{cor1}}

\cortext[cor1]{Corresponding author. \\
\hspace*{1.2em} {\itshape{E-mail address:}} chen\_qiangqiang@163.com (Q. Chen), keyunfeng@stu.kust.edu.cn (Y. Ke), jhlixjtu@163.com (J. Li),lishen@kust.edu.cn (S. Li).}


\address[km1]{Faculty of Science, Kunming University of Science and Technology,
Yunnan 650500, China}

\address[km2]{Data Science Research Center, Kunming University of Science and Technology, Yunnan 650500, China }

\begin{abstract}
Formal Concept Analysis (FCA) is extensively used in knowledge extraction, cognitive concept learning, and data mining. However, its computational demands on large-scale datasets often require outsourcing to external computing services, raising concerns about the leakage of sensitive information.
To address this challenge, we propose a novel approach to enhance data security and privacy in FCA-based computations. Specifically, we introduce a Privacy-preserving Formal Context Analysis (PFCA) framework that combines binary data representation with homomorphic encryption techniques.
This method enables secure and efficient concept construction without revealing private data. Experimental results and security analysis confirm the effectiveness of our approach in preserving privacy while maintaining computational performance. These findings have important implications for privacy-preserving data mining and secure knowledge discovery in large-scale FCA applications.
\end{abstract}


\end{frontmatter}

\parskip=1mm
\section{Introduction}

Formal Concept Analysis (FCA) \cite{1} is a set-theoretic method for modeling conceptual structures in datasets and has been widely studied and applied in domains such as data mining \cite{2}, pattern recognition \cite{3}, conflict analysis \cite{4}, cognitive learning \cite{5}, knowledge discovery \cite{6} and outlier detection \cite{7}.

Despite its wide range of applications, FCA faces computational challenges, especially in the era of exponential data growth. Although many efficient algorithms exist (e.g. batch \cite{batch}, incremental \cite{incremental}, CbO \cite{CbO}, In-Close \cite{In-Close}, Bit-Close \cite{Bit-Close}), further efficiency often requires parallel computing. Efforts include a polynomial-time Galois lattice algorithm \cite{Njiwoua1997}, a parallelized Next-Closure method \cite{Fu2004}, and a divide-and-conquer approach \cite{Kengue2005}. Other implementations include PCbO \cite{PCbO}, FPCbO \cite{FPCbO}, and CUDA-based CbO \cite{cudaCbO}. However, reliance on third-party services poses privacy risks, making it essential to develop secure methods for constructing concept lattices without exposing sensitive data.

Privacy computing \cite{13} aims to enable collaborative analysis without compromising data confidentiality \cite{14,15}. Techniques like federated learning and differential privacy have shown promise in domains like recommendation systems and IoT. Federated learning ensures local data protection via distributed training \cite{10}, though it faces communication overheads in IoT contexts \cite{19,20}. Differential privacy protects data by adding noise while preserving analytical utility \cite{11}, and its integration with blockchain can boost transparency and reduce leakage risks \cite{21,22,23,24}. Additionally, Privacy-Preserving Association Rule Mining (PPARM) has advanced methods to identify sensitive rules and enhance data security \cite{12}.

Sellami et al. \cite{8} introduced FedFCA, a federated learning framework incorporating differential privacy to protect local data in FCA. However, privacy-preserving applications of FCA remain limited. Beyond data confidentiality, FCA must also ensure the integrity and correctness of concept lattice construction under encryption. Unlike conventional privacy-preserving tasks, FCA involves intensive logical operations and may retain fragments of original data, making the construction of secure concepts particularly challenging. Traditional cryptographic methods such as symmetric encryption, asymmetric encryption, block and hybrid encryption, zero-knowledge proofs, hash algorithms, and stream ciphers offer protection during storage or transmission, but require decryption for computation, risking privacy exposure. In addition, many are not suitable for the logical processing demands of FCA. Table \ref{tab3} summarizes the applicability, benefits, and limitations of these techniques. To overcome these challenges, we propose a privacy-preserving FCA method based on homomorphic encryption, which allows operations on encrypted data. Our approach involves two stages: (1) concept processing over encrypted data to generate intermediate results, and (2) selective decryption of non-sensitive results to validate and finalize the concept lattice.

Since homomorphic encryption permits operations on encrypted data without revealing private information, it enables secure data processing.
Homomorphic encryption, first proposed by Rivest et al. \cite{29}, enables secure computations without revealing raw data. The initial schemes supported multiplicative (e.g., RSA) or additive (e.g. Paillier \cite{30}) operations. Gentry’s 2009 breakthrough \cite{31} introduced fully homomorphic encryption (FHE) using ideal lattices, later optimized by Brakerski et al. via the BGV scheme \cite{32}. The CKKS scheme by Cheon et al. \cite{33} further extended FHE to real and complex numbers. These advancements have enabled privacy-preserving collaboration in sectors such as healthcare, transportation, and industry \cite{25,26,27,28}.

\begin{table}[h!]
\centering
\caption{Mainstream encryption technologies and their characteristics.}
\label{tab3}
\begin{tabularx}{\textwidth}{c X X X}
\toprule
\textbf{Encryption Type} & \textbf{Application Scenarios} & \textbf{Advantages} & \textbf{Disadvantages} \\
\midrule
\multirow{2}{*}{Symmetric \cite{35}} & High-speed encryption for large-scale or batch data, e.g., databases or file systems. & Fast processing with low hardware requirements. & Requires secure key distribution and storage; both parties must manage the same secret key. \\
\hline
\multirow{2}{*}{Asymmetric \cite{35}} & Used in small-scale encryption, digital signatures, and secure key exchanges (e.g., TLS/SSL). & Facilitates public key distribution and supports identity authentication. & Computationally intensive; inefficient for large data volumes. \\
\hline
\multirow{2}{*}{Hash Algorithms \cite{36}} & Password storage, integrity verification, and cryptographic signatures. & Fast computation; outputs fixed-length digests ideal for verification. & Vulnerable to collisions in older algorithms (e.g., MD5, SHA-1). \\
\hline
\multirow{2}{*}{Stream Encryption \cite{37}} & Real-time encryption of data streams such as voice or video. & Low latency and high-speed processing. & Legacy algorithms (e.g., RC4) are insecure; robust alternatives like ChaCha20 preferred. \\
\hline
\multirow{2}{*}{Block Encryption \cite{38}} & Used for file/database encryption; often combined with modes (e.g., CBC, GCM). & Supports flexible, secure encryption modes ensuring confidentiality and integrity. & Security varies by mode; ECB may reveal plaintext structure. \\
\hline
\multirow{2}{*}{Hybrid Encryption \cite{39}} & Common in HTTPS/TLS: asymmetric encryption secures keys, symmetric handles data. & Combines asymmetric key exchange with symmetric efficiency. & More complex implementation requiring multi-algorithm coordination. \\
\bottomrule
\end{tabularx}
\end{table}

This paper proposes a homomorphic encryption-based algorithm for secure concept construction in FCA (Formal Concept Analysis) addressing the following questions:

1. Why integrate encryption algorithms into concept construction?

Incorporating encryption algorithms maintains privacy while performing essential operations such as addition and multiplication on ciphertexts. This approach allows FCA to construct concepts as if the data were in plaintext, ensuring concept construction confidentiality.

2. Why choose homomorphic encryption?

Concept construction is a computationally intensive task requiring numerous logical operations on data. When the formal context is encrypted, the conceptual construction operators corresponding to must be performed on ciphertext. Homomorphic encryption ensures that the decrypted result matches the outcome of the same operation on the original data, thus preserving privacy throughout the construction process.

3. What are the innovations of this paper?
\begin{itemize}
\item A privacy-preserving concept construction framework based on homomorphic encryption is proposed, enabling secure logical operations on encrypted formal contexts without revealing sensitive data.
\item A novel, parallelizable two-stage concept construction algorithm is designed, which performs encrypted computation followed by selective decryption of non-sensitive intermediate results, ensuring both efficiency and confidentiality.
\item Comprehensive evaluation and security analysis are conducted, demonstrating the correctness, feasibility, and privacy guarantees of the proposed PFCA method, with experimental validation on real-world UCI datasets.
\end{itemize}

The structure of the rest of this paper is as follows. Section 2 introduces the foundational notions of FCA and homomorphic encryption. Section 3 discusses the notion of privacy in this context, providing a formal definition and related computational methods. Section 4 provides illustrative examples and rigorously demonstrates the correctness of the proposed approach. Finally, Section 5 concludes the paper by summarizing key findings and discussing directions for future research. Finally, Section 6 summarizes the study and highlights prospective directions for future research.

\section{Preliminaries}
In this section, we review some basic notions of formal concept analysis and homomorphic encryption.

\subsection{Formal concept analysis}
\noindent\textbf{Definition 1}. The triplet $\mathcal{F}=(U,A,I)$ is called a formal context where $U=\{o_1,o_2,...,o_m\}$ is a non-empty and finite object set, $A=\{a_1,a_2,...,a_n\}$ represents a non-empty and finite attribute set, and $I: U \times A \rightarrow \{0, 1\}$ is a binary relation. $I(o_i, a_j)=1$ denotes that the object $o_i$ possesses the attribute $a_j$, while $I(o_i, a_j)=0$ denotes that the object $o_i$ does not possess the attribute $a_j$.

\begin{table*}[ht]
	\centering
	\caption{A formal context $\mathcal{F}=(U,A,I)$. \label{tab1}}
	\begin{tabular*}{\textwidth}{@{\extracolsep{\fill}}cccccc@{\extracolsep{\fill}}}
		\toprule
		$U$ & $a_1$ & $a_2$ & $a_3$ & $a_4$ & $a_5$ \\
		\midrule
		$o_1$ & 1 & 0 & 0 & 0 & 0 \\
		$o_2$ & 1 & 1 & 0 & 0 & 0 \\
		$o_3$ & 0 & 0 & 1 & 0 & 1 \\
		$o_4$ & 1 & 1 & 1 & 1 & 0 \\
		\bottomrule
	\end{tabular*}
\end{table*}

\noindent\textbf{Example 1}. Table \ref{tab1} gives an example of formal context where $m=4, n=5$. The set of objects $U = \{ o_1, o_2, \dots, o_4 \}$ represents 4 objects, and the attribute set $A = \{ a_1, a_2, \dots, a_5 \}$ denotes 5 attributes. The number ``1'' represents the object possess this attribute, while number ``0'' indicates the object does not have the attribute. Normally, the formal context that does not have the full row, full column, empty row, and empty column where full means all-1 vector and empty means all-0 vector.

To build concepts from formal context, we need to formalize the intent and extent. First, we define the following operators: $\forall X \subseteq U, B \subseteq A$,
\begin{align}
	f(X) := \{a\in A:\forall x \in X, I(x,a) = 1\},\\
	g(B) := \{x\in U:\forall a \in B, I(x,a) = 1\}.
\end{align}

\noindent\textbf{Definition 2}. Let $\mathcal{F}=(U,A,I)$ be a formal context. $\forall X \subseteq U, B \subseteq A$, if $f(X)=B \land g(B)=X$, the pair $(X,B)$ is a formal concept (concept for short); $X$ is called the extent of concept $(X,B)$, $B$ is called the intent of concept $(X,B)$.

In FCA, a granular concept refers to a specific type of concept that is derived directly from an individual object or attribute within a formal context. Generally speaking, both object-concept and attribute-concept are called granular concept.

\noindent\textbf{Definition 3}. Let $\mathcal{F}=(U,A,I)$ be a formal context, given an object $x \in U$, the object-concept induced by $x$ can be obtained: $(gf(x), f(x))$. Similarly, given an attribute $a \in A$, the attribute-concept induced by $a$ can be obtained: $(g(a), fg(a))$. Granular concepts noted as $G_{fg}$ include both object concept and attribute concept.

\subsection{Hommomorphic encryption}

\noindent\textbf{Definition 4}. Let $\mathcal{E}$ be an encryption function and $\mathcal{D}$ be a decryption function. A fully homomorphic encryption system is an encryption method where the encryption function $\mathcal{E}$ and decryption function $\mathcal{D}$ satisfy the condition that, for any plaintexts $u_1, u_2$ and any binary operation $\odot$, there exists a corresponding ciphertext space operation $\boxdot$ such that
\begin{align}
	\mathcal{D}(\mathcal{E}(u_1) \boxdot \mathcal{E}(u_2)) = u_1 \odot u_2.
\end{align}

This means that the decryption of the result of the ciphertext operation $\boxdot$ on $\mathcal{E}(u_1)$ and $\mathcal{E}(u_2)$ yields the same result as applying the plaintext operation $\odot$ directly on $u_1$ and $u_2$. Based on Definition 4, it is evident that homomorphism is the fundamental property that guarantees the encryption and decryption function preserve the structure of operations. This ensures that operations performed on encrypted data correspond accurately to those on plaintexts. In addition, the binary operation $\odot$ can be any operation such as addition, multiplication, or other algebraic functions. The corresponding operation $\boxdot$ on ciphertexts is constructed to mirror the plaintext operation through encryption and decryption.

\noindent\textbf{Property 1}. The fully homomorphic system has the following three properties:
\begin{enumerate}
	\item Correctness: for all function $F$ which will induce the corresponding homomorphic function $\tilde{F}$ by homomorphic encryption, and all plaintext $u_1, u_2, \cdots, u_m$, the following holds:
	
	\begin{equation}
		\mathcal{D}(\tilde{F}(\mathcal{E}(u_1), \cdots, \mathcal{E}(u_m)) = F(u_1, \cdots, u_m).
	\end{equation}	
	
	\item Semantic security: $\forall u_1, u_2, \mathcal{E}(u_1)\approx \mathcal{E}(u_2)$, which means polynomial indistinguishable, i.e. there is no efficient algorithm that can distinguish between two results, even if $u_1$, $u_2$ are known. This is due to the use of random numbers in the encryption algorithm, even for the same plaintext encryption, the results are not the same.
	
	\item Compactness: for all function $F$, all plaintext $u_1, u_2, \cdots, u_m$, and the responding ciphertexts $c_1, c_2, \cdots, c_m$ the following holds:
	
	\begin{equation}
		\mid F(c_1, c_2, \cdots, c_m) \mid = \textit{poly}(\lambda).
	\end{equation}
	where $\lambda$ is independent of $F$, which means the length of $F(c_1, c_2, \cdots, c_m)$ will not change as $F$ grows bigger.
\end{enumerate}

\noindent\textbf{Property 2}. In a fully homomorphic system, for general addition and multiplication, we can get the corresponding homomorphic addition $\oplus$ and homomorphic multiplication $\otimes$, that is, $\forall a, b \in R$, the following holds:
\begin{gather}
	 a + b = \mathcal{D}(\mathcal{E}(a) \oplus \mathcal{E}(b)), \\
	 a \times b = \mathcal{D}(\mathcal{E}(a) \otimes \mathcal{E}(b)).
\end{gather}

\section{Homomorphic encryption-based concept construction}
In this section, we first introduce how to encrypt the formal context using a fully homomorphic encryption (FHE) method. Then, we construct operators on this basis, lay a mathematical foundation for the concept of privacy protection, formalize its construction conditions, and implement the method through algorithms. In the process, we will explain the encryption method and the calculation process in the encrypted state in detail through examples.

\begin{figure*}[hbtp]
	\centering
	\includegraphics[width=0.80\textwidth]{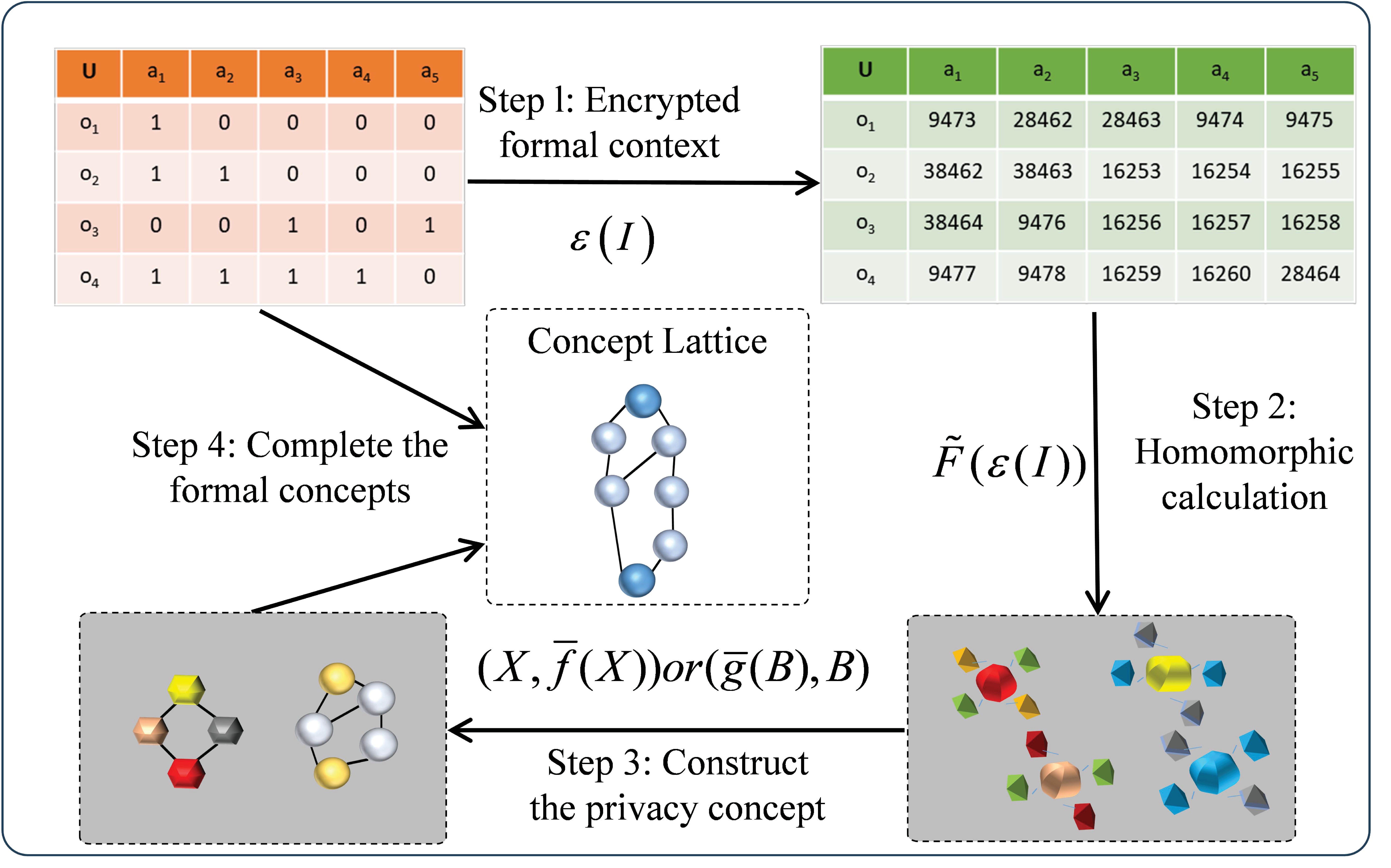}
	\caption{The diagram of the privacy computing for concept.}
\end{figure*}

\subsection{Privacy-preserving formal context}

\begin{table*}[ht]
	\centering
	\caption{A privacy formal context $\mathcal{F'}=(U,A,I^*)$ responding to Table 1. \label{tab2}}
	\begin{tabular*}{\textwidth}{@{\extracolsep{\fill}}cccccc@{\extracolsep{\fill}}}
		\toprule
		$U$ & $a_1$ & $a_2$ & $a_3$ & $a_4$ & $a_5$ \\
		\midrule
		$o_1$ & 9473  & 28462  & 28463  & 9474  & 9475  \\
		$o_2$ & 38462 & 38463  & 16253  & 16254  & 16255 \\
		$o_3$ & 38464 & 9476   & 16256  & 16257  & 16258 \\
		$o_4$ & 9477  & 9478   & 16259  & 16260  & 28464 \\
		\bottomrule
	\end{tabular*}
\end{table*}

\noindent\textbf{Definition 5}.
Let $\mathcal{F'}=(U, A, I^*)$ be a privacy formal context, where $I^*$ is the encrypted form of the original relation $I$. Specifically, $I^*$ is an encrypted $m \times n$ matrix that preserves the structural information of the formal context while ensuring data privacy. The encryption function $\mathcal{E}$ and the decryption function $\mathcal{D}$ are defined such that applying the encryption function to $I$ yields $I^* = \mathcal{E}(I)$.

The use of FHE ensures that operations can be performed on $I^*$ without exposing or compromising the privacy of the underlying data. Definition 5 aligns with Step 1 illustrated in Figure 1.

\noindent\textbf{Example 2}
We encrypt Table \ref{tab1} according to Definition 5 using FHE, where $U = \{ o_1, o_2, \dots, o_4 \}$ and $A = \{ a_1, a_2, \dots, a_5 \}$ represent the encrypted object and attribute sets, respectively. Table \ref{tab2} presents a privacy formal context after applying encryption function $\mathcal{E}$, where $\mathcal{E}$ using the widely adopted homomorphic encryption library.

\noindent\textbf{Definition 6}.
Let $\mathbf{u}$ and $\mathbf{v}$ be two vectors of length $n$. We define a operation $\tilde{\otimes}$, derived from the homomorphic multiplication function $\otimes$, which returns a vector $\mathbf{w}$ of length $n$, the same as the input vector. The operation is defined as follows:
\begin{align}
	\mathbf{w} = \mathbf{u} \tilde{\otimes} \mathbf{v},
\end{align}
where $w^i = u^i \otimes v^i$ ($i = 1, 2, \dots, n$), with $i$ indicating the $i$-th component of the vectors.

\noindent\textbf{Definition 7}.
Let $\mathbf{u} = \{\mathbf{u_1}, \mathbf{u_2}, \dots, \mathbf{u_m}\}$ be a set of $m$ vectors, each of dimension $n$, where $m \geq 2$. Each $\mathbf{u_j}$ is defined as $\mathbf{u_j} = (u_j^1, u_j^2, \dots, u_j^n)$ ($j = 1, 2, \dots, m$). The function $P(\mathbf{u})$ is expressed as:
\begin{align}
	u = P(\mathbf{u}) = \mathbf{u_1} \tilde{\otimes} \mathbf{u_2} \tilde{\otimes} \dots \tilde{\otimes} \mathbf{u_m},
\end{align}
where $u$ is a vector of length $n$, with each element ${u}^i$ defined as ${u}^i = {u}_1^i \otimes {u}_2^i \otimes \dots \otimes {u}_m^i$.

\noindent\textbf{Definition 8}.
Let $\mathbf{u} = \{{u^1}, {u^2}, \dots, {u^n}\}$ be a vector of length $n$, we define the function $S$ as the sum of all the elements of $\mathbf{u}$ as follows:
\begin{align}
	s = S(\mathbf{u}) = u^1 \oplus u^2 \oplus \dots \oplus u^n,
\end{align}
where $\oplus$ denotes the homomorphic addition operator.

\noindent\textbf{Definition 9}.
Let$ P $ and $ S $ be two functions. We define $ F $ as the composition of $ P $ and $ S $, expressed as $ F = S \circ P $, where $ F $ maps a set of $ m $ vectors with length $n$ to an integer. In the formal context $\mathcal{F'}$, the plaintext functions $ P $ and $ S $ according to the homomorphic ciphertext functions $ \tilde{P} $ and $ \tilde{S} $, respectively. Therefore, the function $ F $ has a corresponding homomorphic counterpart $ \tilde{F} $.

By encrypting the plaintext vectors $\mathbf{u_1}, \mathbf{u_2}, \dots, \mathbf{u_m}$, we obtain the corresponding ciphertext vectors $\mathbf{c_1}, \mathbf{c_2}, \dots, \mathbf{c_m}$, where $\mathbf{c_j} = \mathcal{E}(\mathbf{u_j})$. Subsequently, we apply the homomorphic composition $ \tilde{S} \circ \tilde{P}$ to the ciphertext vectors, yielding $ \tilde{S} \circ \tilde{P}(\mathbf{c_1}, \mathbf{c_2}, \dots, \mathbf{c_m}) = \tilde{S}(\tilde{P}(\mathbf{c_1}, \mathbf{c_2}, \dots, \mathbf{c_m}))$. By the properties of the FHE system, we know that $ \mathcal{D}(\tilde{S} \circ \tilde{P}(\mathbf{c_1}, \mathbf{c_2}, \dots, \mathbf{c_m})) = S(P(\mathbf{u_1}, \mathbf{u_2}, \dots, \mathbf{u_m}))$, where $\mathcal{D}$ is the decryption function, and $\mathcal{E}$ is the encryption function.

We define the computation of the sum of the elements of vector $\mathbf{c_1}$ using homomorphic addition $\oplus$ as follows:
\begin{align}
	s = c_1^1 \oplus c_1^2 \oplus \cdots \oplus c_1^n,
\end{align}
where $s$ represents the homomorphic sum of all elements of $\mathbf{c_1}$.

By encrypting the plaintext vectors, performing homomorphic operations on the ciphertexts, and then decrypting the results, the system enables secure computation of plaintext functions while maintaining data privacy. Definitions 6, 7, 8, and 9 collectively outline the operations carried out in Step 2 of Figure 1, offering a thorough theoretical foundation for this stage.

\noindent\textbf{Definition 10}.
Let $\mathbf{c_1} = (c_1^1, c_1^2, \ldots, c_1^n)$ and $\mathbf{c_2} = (c_2^1, c_2^2, \ldots, c_2^n)$ be two homomorphically encrypted vectors. The homomorphic comparison function $\alpha(\mathbf{c_1}, \mathbf{c_2})$ is defined as follows:
\begin{align}
\alpha(\mathbf{c_1}, \mathbf{c_2}) = \left( s - \tilde{S} \circ \tilde{P}(\mathbf{c_1}, \mathbf{c_2}) \right) \otimes \mathcal{E}(r),
\end{align}
where $\mathcal{E}(r)$ is the homomorphic encryption of a random value $r$.

If the result $\mathcal{D}(\alpha(\mathbf{c_1}, \mathbf{c_2}))$ is $0$, the two vectors are the same, otherwise we know nothing but they are not the same. For convenience, if the result is 0, it returns true, otherwise, it returns false.

\subsection{Homomorphic encryption-based concept construction algorithm}
Let $ f $ be the operator in a formal context $\mathcal{F}$ that retrieves the common attributes shared by the object set $ A $. In the corresponding encrypted formal context $\mathcal{F'}$, for simplicity, we treat $ \mathbf{o_i} $ as the $ i $-th row of $ I^* $. The operator $ \Tilde{f}(X)$ according to $\mathcal{F'}$ is defined as follow:
\begin{align}
	 \Tilde{f}(X) = \mathcal{D}(\tilde{S} \circ \tilde{P}(\mathbf{o_{i_1}}, \mathbf{o_{i_2}}, \dots, \mathbf{o_{i_k}})),
\end{align}
where $ X $ is an index set with $ X \subseteq \{1, 2, \dots, m\} $. Similarly, we define $ \Tilde{g}(B) $ as follow:
\begin{align}
	 \Tilde{g}(B) = \mathcal{D}(\tilde{S} \circ \tilde{P}(\mathbf{a_{j_1}}, \mathbf{a_{j_2}}, \dots, \mathbf{a_{j_l}})),
\end{align}
where $ B $ is an index set with $ B \subseteq \{1, 2, \dots, n\} $. Here, $ \mathbf{a_j} $ represents the $ j $-th column of $ I^* $.

\noindent\textbf{Property 3}.
For any sets $ X_1, X_2 \subseteq \{1, 2, \dots, m\} $, we have $\Tilde{f}(X_1 \cup X_2) = \Tilde{f}(X_2 \cup X_1)$. In particular, when $ X_1 = X_2 = X $, it follows that $\Tilde{f}(X) \tilde{\otimes} \Tilde{f}(X) = \Tilde{f}(X).$

\noindent\textbf{Proof.} Since the union operation $ X_1 \cup X_2 $ is commutative, the homomorphic operator $ \tilde{P} $ applied to rows indexed by $ X_1 \cup X_2 $ is symmetric. Thus, $\Tilde{f}(X_1 \cup X_2) = \mathcal{D}(\tilde{S} \circ \tilde{P}(\mathbf{o}_{i_1}, \mathbf{o}_{i_2}, \dots)) = \Tilde{f}(X_2 \cup X_1)$. When $ X_1 = X_2 = X$, according to Definition 7, repeatedly applying $ \tilde{\otimes}$ to $ \Tilde{f}(X)$ does not introduce any new components. In other words, $\Tilde{f}(X)$ remains unchanged, i.e.,$\Tilde{f}(X) = \mathcal{D}(\tilde{S} \circ \tilde{P}(\mathbf{o}_{i_1}, \mathbf{o}_{i_2}, \dots)).$ Thus, applying $ \tilde{\otimes}$ to $ \Tilde{f}(X) $ with itself yields $\Tilde{f}(X) \tilde{\otimes} \Tilde{f}(X) = \Tilde{f}(X)$, because the homomorphic product is idempotent when applied to identical inputs. \hfill $\blacksquare$\par

\noindent\textbf{Example 3}
Say we now have $X=\{1, 2\}$, the $\mathbf{r_1}$ and $\mathbf{r_2}$ correspond to the first row and second row in Table 1, and the $\mathbf{o_1}$ and $\mathbf{o_2}$ correspond to the first row and second row in Table 2. That is, $\mathbf{o_1} = \mathcal{E}(\mathbf{r_1})$, and $\mathbf{o_2} = \mathcal{E}(\mathbf{r_2})$. Then $\Tilde{f}(X)= \mathcal{D}(\tilde{S} \circ \tilde{P}(\mathbf{o_1}, \mathbf{o_2})) \stackrel{*}{=} S (P (\mathbf{r_1}, \mathbf{r_2})) = 1 \times 1 + 0 \times 1 + 0 \times 0 + 0 \times 0 + 0 \times 0 = 1$. The equal * is induced by the correctness property, and just to make it clear how the $\Tilde{f}$ work. The actual operation that happens is $\Tilde{f}(X) = \mathcal{D} (9473 \otimes 38462 \oplus 28462 \otimes 38463 \oplus 28463 \otimes 16253 \oplus 9474 \otimes 16254 \oplus 9475 \otimes 16255) = 1$. Similarly, assuming we have $B=\{1, 3\}$, the $\mathbf{s_1}$ and $\mathbf{s_3}$ correspond to the first column and third column in Table 1, and the $\mathbf{a_1}$ and $\mathbf{a_3}$ correspond to the first column and third column in Table 2. That is, $\mathbf{a_1} = \mathcal{E}(\mathbf{s_1})$, and $\mathbf{a_3} = \mathcal{E}(\mathbf{s_3})$. Then $\Tilde{g}(B) = \mathcal{D}(\tilde{S} \circ \tilde{P}(\mathbf{a_1}, \mathbf{a_3})) \stackrel{*}{=} S(P(\mathbf{s_1}, \mathbf{s_3})) = 1 \times 0 + 1 \times 0 + 0 \times 1 + 1 \times 1 = 1$, the equal * is also induced by the correctness property. The actual operation that happens is $\Tilde{g}(B) = \mathcal{D} (9473 \otimes 28463 \oplus 38462 \otimes 16253 \oplus 38464 \otimes 16256 \oplus 9477 \otimes 9478) = 1$.

Unlike the normal concept, now we don't have the whole information which tells us whether the concept $(X,Y)$ satisfying $f(X)=Y\land g(Y)=X$. Instead, we now have two ways to build the concept, by operator $\tilde{f}$ or $\tilde{g}$, and the concepts are termed as $\tilde{f}$-induced privacy-preserving concept and $\tilde{g}$-induced privacy-preserving concept, respectively.

\noindent\textbf{Definition 11}. Let $(U, A, I^*)$ be the privacy formal context, where $X \subset U$, $B \subset A$, and $\tilde{f}$ and $\tilde{g}$ are operators as defined. If $\tilde{f}(X) > 0$, and for all $x' \in U$ such that $x' \notin X$, $\tilde{f}(X \cup \{x'\}) < \tilde{f}(X)$, then $(X, \tilde{f}(X))$ is a $\tilde{f}$-induced privacy-preserving concept. Similarly, if $\tilde{g}(B) > 0$, and for all $a' \in A$ such that $a' \notin B$, $\tilde{g}(B \cup \{a'\}) < \tilde{g}(B)$, then $(\tilde{g}(B), B)$ is a $\tilde{g}$-induced privacy-preserving concept.

To build $\tilde{f}$-induced privacy-preserving concept, we map the $\tilde{f}$ to every possible extent $X$ where $\mid X\mid > 1$, we will have three situations:

(1) $\tilde{f}(X)=0$, then $(X, \tilde{f}(X))$ will not be a privacy-preserving concept;

(2) if $\tilde{f}(X)>0$, and $\forall x' \in U \land x' \notin X,\tilde{f}(X\cup \{x'\})<\tilde{f}(X)$, then $X$ is the extent, and the corresponding privacy-preserving concept is $(X, \tilde{f}(X))$.

(3) if $\tilde{f}(X)>0$, but $\exists x' \in U \land x' \notin X, \tilde{f}(X\cup \{x'\})=\tilde{f}(X)$, then $(X, \tilde{f}(X))$ is not a pricacy concept.

It should be noted that there are three special cases that $\tilde{f}$ can not work: $\mid X\mid = 1$, $\mid X \mid = 0$, and $X = U$. For extent $X$ where $\mid X\mid = 1$, we have a homomorphic compare function $\alpha$ which takes two encrypted vectors and return a number. The object in $X$ is denoted as $X$, the any other object in $U / X$ is denoted as $x'$. Without confusion, we're going to treat $\mathbf{X}$ as the $X$-th row in $I^*$ which is a vector. It was the same for $\mathbf{x'}$. If $\mathcal{D}(\alpha(\mathbf{X}, \mathbf{x'}))$ is true, we add the corresponding object into the current extent. And for the special case where $\mid X \mid = 0$, we have to use the dual operator $\tilde{g}$ and compute $\tilde{g}(G)$, if $\tilde{g}(G) = 0$, then $(\phi, 0)$ forms a privacy-preserving concept. Last but not least, for the final special case where $X = U$, if $\tilde{f}(X) = 0$, the $(U, 0)$ forms a privacy-preserving concept.

The construction of $\tilde{f}$-induced privacy-preserving concept is based on determining whether a range $X$ qualifies as a privacy-preserving concept by analyzing the size of $X$ and the variation characteristics of $\tilde{f}(X)$. In special cases, homomorphic comparison function $\alpha$ or dual operator $\tilde{g}$ is used as a supplement, thereby fully defining the rule system for privacy-preserving concept. Algorithm 1 presents the construction process of $\tilde{f}$-induced privacy-preserving concept, with a time complexity of $O(2^m \cdot (f_c + m))$.

\begin{algorithm}[htb]
	\caption{Using the $\tilde{f}$ operator to construct privacy-preserving concept}
	\label{alg:algorithm1}
	\begin{algorithmic}[1]
		\REQUIRE Privacy formal context $(U, A, I^*)$
		\ENSURE $\tilde{f}$-induced privacy-preserving concept
		
		\FOR{each $X$ in $2^U$}
		\IF{$|X| > 1$}
		\IF{$\tilde{f}(X) = 0$}
		\STATE $(X, \tilde{f}(X))$ is not a privacy-preserving concept
		\ELSIF{$\tilde{f}(X) > 0$ \AND $\forall x' \in U, x' \notin X, \tilde{f}(X \cup \{x'\}) < \tilde{f}(X)$}
		\STATE $(X, \tilde{f}(X))$ is a privacy-preserving concept
		\ELSIF{$\tilde{f}(X) > 0$ \AND $\exists x' \in U, x' \notin X, \tilde{f}(X \cup \{x'\}) = \tilde{f}(X)$}
		\STATE $(X, \tilde{f}(X))$ is not a privacy-preserving concept
		\ENDIF
		\ELSIF{$|X| = 1$}
		\STATE Initialize privacy-preserving concept $(\mathrm{Extent}, p) \leftarrow (\phi, 0)$
		\FOR{each $x' \in U$ \AND $x' \neq X$}
		\IF{$\mathcal{D}(\alpha(\mathbf{X}, \mathbf{x'}))$ is True}
		\STATE Update privacy-preserving concept: $(\mathrm{Extent} \cup x', p + 1)$
		\ENDIF
		\ENDFOR
		\ELSIF{$|X| = 0$}
		\IF{$\tilde{g}(A) = 0$}
		\STATE $(\phi, 0)$ is a privacy-preserving concept
		\ENDIF
		\ELSIF{$|X| = m$}
		\IF{$\tilde{f}(U) = 0$}
		\STATE $(U, 0)$ is a privacy-preserving concept
		\ENDIF
		\ENDIF
		\ENDFOR
	\end{algorithmic}
\end{algorithm}

To build $\tilde{g}$-induced privacy-preserving concept, we can map the $\tilde{g}$ to every possible intent $B$ where $\mid B\mid > 1$, we will also have three situations:

(1) $\tilde{g}(B)=0$, then $(\tilde{g}(B), B)$ will not be a concept;

(2) if $\tilde{g}(B)>0$, and $\forall a' \in A \land a' \notin B,\tilde{g}(B\cup \{a'\})<\tilde{g}(B)$, then $B$ is the intent, and the corresponding privacy-preserving concept is $(\tilde{g}(B), B)$.

(3) if $\tilde{g}(B)>0$, but $\exists a' \in A \land a' \notin B, \tilde{g}(B\cup \{a'\})=\tilde{g}(B)$, then $(\tilde{g}(B), B)$ is not a pricacy concept, we call it pseudo-concept.

The same as $\tilde{f}$, there are three special cases that $\tilde{g}$ can not work: $\mid B\mid = 1$, $\mid B \mid = 0$, and $B = A$. For intent $B$ where $\mid B \mid = 1$. The attribute in $B$ is denoted as $B$, the any other object in $A / B$ is denoted as $b'$. Without confusion, we're going to treat $\mathbf{B}$ as the $B$-th column in $I^*$ which is a vector. It was the same for $\mathbf{b'}$. If $\mathcal{D}(\tilde{c}(\mathbf{B},\mathbf{b'}))$ is true, we add the corresponding attribute into the current intent. And for the special case where $\mid B \mid = 0$, we have to use the dual operator $\tilde{f}$ and compute $\tilde{f}(U)$, if $\tilde{f}(U) = 0$, then $(0, \phi)$ forms a privacy-preserving concept. Finally, for the last special case where $B = A$, if $\tilde{g}(B) = 0$, the $(0, A)$ forms a privacy-preserving concept.

Algorithm 2 presents the construction process of $\tilde{g}$-induced privacy-preserving concept, with a time complexity of $O(2^n \cdot (g_c + n))$. $\tilde{f}$-induced and $\tilde{g}$-induced privacy-preserving concept are collectively termed as privacy-preserving concept. Definitions 11 correspond to the operations described in Step 3 of Figure 1.

\begin{algorithm}[htb]
	\caption{Using the $\tilde{g}$ operator to construct privacy-preserving concept}
	\label{alg:algorithm2}
	\begin{algorithmic}[1]
		\REQUIRE Privacy formal context $(U, A, I^*)$
		\ENSURE $\tilde{g}$-induced privacy-preserving concept
		
		\FOR{each $B$ in $2^A$}
		\IF{$|B| > 1$}
		\IF{$\tilde{g}(B) = 0$}
		\STATE $(\tilde{g}(B), B)$ is not a privacy-preserving concept
		\ELSIF{$\tilde{g}(B) > 0$ \AND $\forall b' \in A, b' \notin B, \tilde{g}(B \cup \{b'\}) < \tilde{g}(B)$}
		\STATE $(\tilde{g}(B), B)$ is a privacy-preserving concept
		\ELSIF{$\tilde{g}(B) > 0$ \AND $\exists b' \in A, b' \notin B, \tilde{g}(B \cup \{b'\}) = \tilde{g}(B)$}
		\STATE $(\tilde{g}(B), B)$ is not a privacy-preserving concept
		\ENDIF
		\ELSIF{$|B| = 1$}
		\STATE Initialize privacy-preserving concept $(q, \mathrm{Intent}) \leftarrow (\phi, 0)$
		\FOR{each $b' \in A$ \AND $b' \neq B$}
		\IF{$\mathcal{D}(\alpha(\mathbf{B}, \mathbf{b'}))$ is True}
		\STATE Update privacy-preserving concept: $(q + 1, \mathrm{Intent} \cup b')$
		\ENDIF
		\ENDFOR
		\ELSIF{$|B| = 0$}
		\IF{$\tilde{g}(B) = 0$}
		\STATE $(0, \phi)$ is a privacy-preserving concept
		\ENDIF
		\ELSIF{$|B| = n$}
		\IF{$\tilde{g}(B) = 0$}
		\STATE $(0, A)$ is a privacy-preserving concept
		\ENDIF
		\ENDIF
		\ENDFOR
	\end{algorithmic}
\end{algorithm}

\noindent\textbf{Example 4} To Table 1, we can get the ordinary formal concept. Since there are fewer objects than attributes, it is more convenient to enumerate all the elements of the power set $P(U)$. So, we have 7 privacy-preserving concepts in total, which are $(\phi, A)$, $(\{o_3\}, \{a_3, a_5\})$, $(\{o_4\}, \{a_1, a_2, a_3, a_4\})$, $(\{o_2, o_4\}, \{a_1, a_2\})$, $(\{o_3, o_4\}, \{a_3\})$, $(\{o_1, o_2, o_4\}, \{a_1\})$, and $(U, \phi)$. Next, we can get the privacy formal privacy-preserving concepts for the encrypted formal context which is shown as Table 2. Because privacy-preserving concept can be induced by operator $\tilde{f}$ or $\tilde{g}$, we first introduce how to form privacy-preserving concept by $\tilde{f}$ as Figure 2.

\begin{figure*}[h]
	\centering
	\includegraphics[width=1\textwidth]{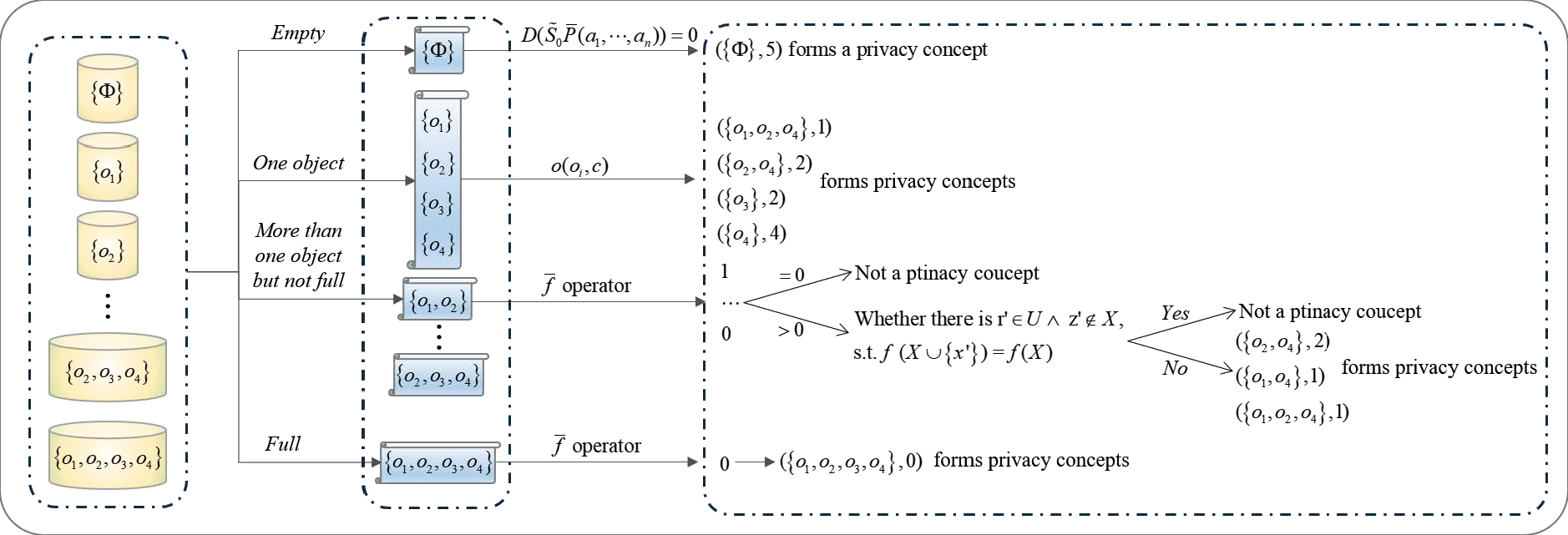}
	\caption{Generate the privacy-preserving concept by $\tilde{f}$ operator.}
	\label{fig1}
\end{figure*}

To conclude, we also have 7 privacy-preserving concept in total, which are $(\phi, 5)$, $(\{o_3\}, 2)$, $(\{o_4\}, 4)$, $(\{o_2, o_4\}, 2)$, $(\{o_3, o_4\}, 1)$, $(\{o_1, o_2, o_4\}, 1)$ and $(U, 0)$. Similarly, we can get 7 privacy-preserving concept by $\tilde{g}$ which are $(0, A)$, $(1, \{a_3, a_5\})$, $(1, \{a_1, a_2, a_3, a_4\})$, $(2, \{a_1, a_2\})$, $(2, \{a_3\})$, $(3, \{a_1\})$ and $(4, \phi)$.

\noindent\textbf{Definition 12}. Let the privacy formal concept be $(X, \tilde{f}(X))$ or $(\tilde{g}(B), B)$, if we have the unencrypted matrix $I$, we can easily get the corresponding formal concept $(X, B)$ where $f(X)=B \land g(B)=X$ by looking up the formal context.

Definition 12 describes the relationship between the privacy formal concept $(X, \tilde{f}(X))$ or $(\tilde{g}(B), B)$ and the formal concept $(X, B)$: through the unencrypted formal context matrix $I$, the formal concept $(X, B)$ can be easily restored, where $f(X) = B$ and $g(B) = X$. This corresponds to Step 4 illustrated in Figure 1.

\noindent\textbf{Example 5}. By Example 4, we obtain the privacy-preserving concept, and in the following we will obtain the corresponding formal concept by decryption. To the $\tilde{f}$-induced privacy-preserving concept, which are $(\phi, n)$, $(\{o_3\}, 2)$, $(\{o_4\}, 4)$, $(\{o_2, o_4\}, 2)$, $(\{o_3, o_4\}, 1)$, $(\{o_1, o_2, o_4\}, 1)$, and $(U, 0)$, we can easily get the formal concept by looking up Table 1, as follows: $(\phi, A)$, $(\{3\}, \{a_3, a_5\})$, $(\{4\}, \{a_1, a_2, a_3, a_4\})$, $(\{2, 4\}, \{a_1, a_2\})$, $(\{3, 4\}, \{a_3\})$, $(\{1, 2, 4\}, \{a_1\})$, and $(U, \phi)$. Similarly, for the $\tilde{g}$-induced privacy-preserving concept, we will get the same result: $(\phi, A)$, $(\{3\}, \{a_3, a_5\})$, $(\{4\}, \{a_1, a_2, a_3, a_4\})$, $(\{2, 4\}, \{a_1, a_2\})$, $(\{3, 4\}, \{a_3\})$, $(\{1, 2, 4\}, \{a_1\})$, and $(U, \phi)$. It is easy to find that the formal concepts obtained by the above two methods are consistent with the original ones.

\section{Analysis of FHE methods based on PFCA}
In this section, we analyze the security of the proposed PFCA from the perspectives of correctness, privacy protection. Additionally, it emphasizes the necessity of homomorphic encryption within the privacy protection framework.

\subsection{The correctness and security of homomorphic encryption}

\noindent\textbf{Theorem 1}
Let $ C_{FCA}$ denote the number of elements in the set of formal concepts generated by FCA, and $ C_{PFCA} $ indicate the total number of formal concepts obtained through PFCA. If the cloud servers faithfully execute the homomorphic computation, then the following conclusion holds:
\begin{align}
	C_{PFCA} = C_{FCA}.
\end{align}
\noindent\textbf{Proof.}
$\Rightarrow$) {Let $\mathcal{F}=(U,A,I)$ be a formal context}. For $X \subseteq U, B \subseteq A$, if $f(X)=B$ and $g(B)=X$, the pair $(X,B)$ is a formal concept. Now, with the Definitions 12 we know that the concept $(\left| X \right|, B)$ will still be a privacy-preserving concept, because operator $f$, $g$ corresponds to the biggest set. That is, $\tilde{f}(X)$ will be the biggest number and there will be no such $x' \in U \land x' \notin X,\tilde{f}(X\cup \{x'\})=\tilde{f}(X)$. Hence, the number of formal concepts is equal to the number of privacy-preserving concepts.

$\Leftarrow$) if $\tilde{f}(X)>0$, and $\forall x' \in U \land x' \notin X,\tilde{f}(X\cup \{x'\})<\tilde{f}(X)$, which means $X$ forms a biggest set, i.e. $g(f(X)) = X$. Otherwise, there will be another $x' \in U /X$ since that $\tilde{f}(X\cup \{x'\}) = \tilde{f}(X)$. That would against the assumption. \hfill $\blacksquare$\par

As shown in Figure 1, the third-party computing platform can only access encrypted data and perform corresponding calculations in steps 2 and 3. The homomorphic encryption method ensures that the encryption state is maintained during the data access and calculation process. This ensures the correctness and security of conceptual cognitive computing.

\noindent\textbf{Theorem 2}
Let $\mathbf{u} = (u^1, u^2, \dots, u^n)$ be a plaintext vector encrypted as $\mathbf{c} = (c^1, c^2, \dots, c^n)$ and $\tilde{S}$ denote the homomorphic summation operation in the ciphertext space. For any adversary $\mathcal{A}$, given the aggregated ciphertext $\mathbf{c_s} = \tilde{S}(\mathbf{c}) = c^1 \oplus c^2 \oplus \dots \oplus c^n$, it is computationally infeasible to deduce the plaintext vector $\mathbf{u}$. Formally,
\begin{align}
	\mathcal{A}(\mathbf{c_s}) \not\Rightarrow \mathbf{u}.
\end{align}

\noindent\textbf{Proof.}
According to the properties of homomorphic encryption, for each element $u^i \in \mathbf{u}$, its ciphertext $c^i = \mathcal{E}(u^i)$ satisfies the indistinguishable property:
\begin{align}
	\forall u^i, v^i \in \mathbb{M}, \quad \mathcal{A}(\mathcal{E}(u^i)) \approx \mathcal{A}(\mathcal{E}(v^i)),
\end{align}
where $\mathbb{M}$ denotes the plaintext space, and $\mathcal{A}$ represents a polynomial-time adversary.
As a result, the adversary cannot deduce the plaintext $u^i$ from the ciphertext $c^i$. In the ciphertext space, the homomorphic summation operation is defined as:
\begin{align}
	\mathbf{c_s} = \tilde{S}(\mathbf{c}) = c^1 \oplus c^2 \oplus \dots \oplus c^n.
\end{align}
Since $\oplus$ is an operation within the encrypted domain, the aggregated result $\mathbf{c_s}$ does not disclose any individual plaintext $u^i$.
Specifically, the adversary can only observe the aggregated ciphertext $\mathbf{c_s}$. However, due to the indistinguishability property of $\mathcal{E}$, it remains infeasible to infer the plaintext vector $\mathbf{u}$. The decryption operation $\mathcal{D}$ satisfies the following:
\begin{align}
	\mathcal{D}(\mathbf{c_s}) = S(\mathbf{u}) = u^1 \oplus u^2 \oplus \dots \oplus u^n.
\end{align}
This implies that decryption reveals only the summation result $s$ rather than the individual elements of $\mathbf{u}$. Even if the adversary obtains $s$, they cannot deduce the specific values of $u^i$. In conclusion, the adversary cannot infer any information about the original plaintext vector $\mathbf{u}$ from the ciphertext summation result $\mathbf{c_s}$. \hfill $\blacksquare$\par

\begin{figure*}[h]
	\centering
	\includegraphics[width=1\textwidth]{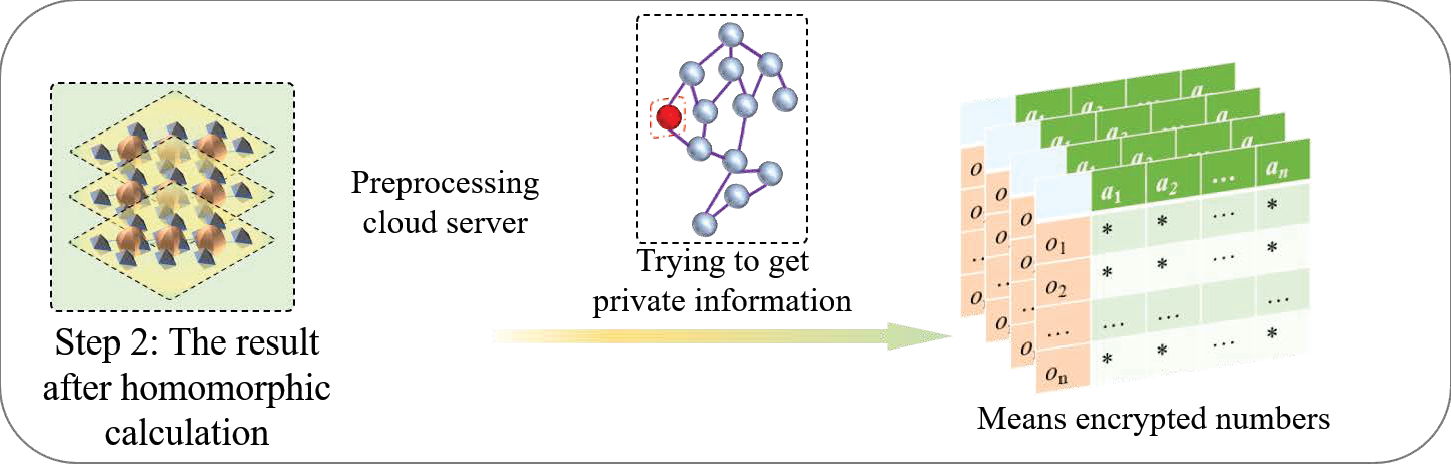}
	\caption{The security analysis of the Step 2.}
	\label{fig1}
\end{figure*}

According to \cite{22}, the CKKS method achieves IND-CPA security. Specifically, based on the IND-CPA property of symmetric encryption, for any two plaintext vectors $ u $ and $ v $, and their corresponding ciphertexts $ \mathcal{E}(u) $ and $ \mathcal{E}(v) $, if a polynomial-time adversary $ \mathcal{A} $ satisfies $ \mathcal{A}(\mathcal{E}(u)) \approx \mathcal{A}(\mathcal{E}(v))$, the adversary cannot distinguish between the ciphertexts of $ u$ and $ v $.

In Step 2, the security of our method relies on the strong properties of the underlying homomorphic encryption. As shown in Figure 3, homomorphic encryption ensures that operations on encrypted data reveal no information about the corresponding plaintext. Even if a third-party observes the computation and inspects the resulting encrypted outputs, the inherent security of the homomorphic encryption prevents any inference of meaningful information about the original data.

\begin{figure*}[h]
	\centering
	\includegraphics[width=1\textwidth]{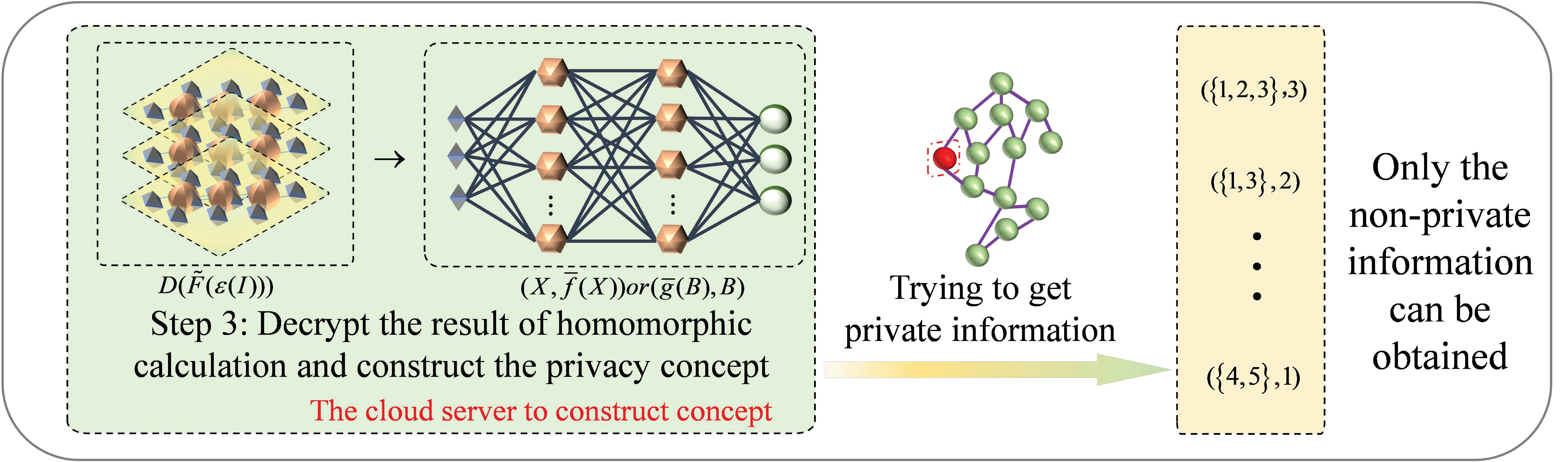}
	\caption{The security analysis of the Step 3.}
	\label{fig1}
\end{figure*}

In Step 3 of the proposed privacy-preserving framework, the intrinsic properties of homomorphic encryption and their implications for privacy protection are highlighted. As depicted in Figure 4, while third-party entities may decrypt the outputs of homomorphic computations, they remain incapable of deducing the original private information of users. This is because operations performed on encrypted data inherently obscure sensitive details. The outputs, treated as intermediate computational results, remain encrypted and abstracted, granting access only to generalized information, such as shared attributes among objects. This approach robustly safeguards user privacy within the cryptographic security boundaries of the homomorphic encryption method, establishing a secure and reliable platform for sensitive data analysis.

\subsection{The necessity of homomorphic encryption}

In the era of deep integration between digitization and networking, encryption technology has become a core means of ensuring information security. Whether for personal privacy protection or safeguarding sensitive data for enterprises and government agencies, encryption technology is indispensable. From symmetric and asymmetric encryption to hash algorithms and digital signatures, and even homomorphic encryption, each technology has distinct characteristics and application scenarios. To effectively select and apply encryption technologies, one must comprehensively consider factors such as performance, security strength, and implementation complexity. With the increasing demands for data privacy and security, especially in sensitive industries such as cloud computing, healthcare, and finance, the ability to perform effective computations without exposing raw data has become a critical issue. 

In contrast to traditional methods, homomorphic encryption overcomes this limitation by allowing specific computational operations to be performed directly on encrypted data while ensuring that the results remain encrypted. This capability enables homomorphic encryption to support complex computational tasks, such as FCA, without exposing the underlying raw data. Under the protection of homomorphic encryption, sensitive data can be processed directly in its encrypted form, enabling FCA algorithms to operate on encrypted data and construct concept lattices without requiring decryption.

The distinguishing feature of homomorphic encryption lies in its ability to perform basic operations, such as addition and multiplication, directly on ciphertext. These operations are fundamental for constructing concept lattices in FCA. By leveraging homomorphic encryption, it is possible to replicate the computational environment of plaintext processing within the encrypted domain. This approach not only ensures data analysis accuracy but also preserves privacy protection, providing a secure and innovative solution for analyzing and processing encrypted data.

\section{Experiments}
In this section, we analyze our method through experiments. To evaluate the feasibility and security of the proposed method, a series of experiments were conducted on a Linux operating system equipped with an AMD EPYC 9354 32-Core CPU running at 3.25 GHz and 32 GB of RAM. The torus-based fully homomorphic encryption method was employed for computational performance, implemented using the widely adopted homomorphic encryption library. The datasets selected for the experiments are from the UCI Machine Learning Repository. Seven representative datasets—Seeds, Mushroom, Nursery, Letter, Appliances Energy Prediction, Adult, Relation Network, and Health Indicators—were selected from the UCI database due to their diverse characteristics and relevance to our analysis. The information about these datasets is presented in Table \ref{tab4}, where "-" indicates that the feature description is not available on UCI.

\begin{table*}[ht]
	\centering
	\caption{Detailed information of datasets.}\label{tab4}
	\begin{tabular*}{\textwidth}{@{\extracolsep{\fill}}lllll@{\extracolsep{\fill}}}
		\toprule
		ID & Dataset & Instances & Features & Size (MB) \\
		\midrule
		1  & Mushroom                     & 8,124    	& 22 (Binary, Categorical)				& 1.83      \\
		2  & Nursery                      & 12,960   	& 8 (Categorical)				        & 0.82      \\
		3  & Appliances Energy Prediction & 19,735   	& 28 (Real)			                    & 11.42     \\
		4  & Letter                       & 20,000   	& 17 (Integer, Categorical)				& 10.70     \\
		5  & Adult                        & 48,842   	& 15 (Integer, Binary, Categorical)		& 12.40     \\
		6  & Relation Network             & 53,413   	& -  (Integer, Real)				    & 7.00      \\
		7  & Health Indicators            & 253,680  	& 21 (Categorical, Integer)				& 21.69     \\
		\bottomrule
	\end{tabular*}
\end{table*}

However, these datasets present several challenges, including multi-valued attributes, real-valued attributes, and missing values. After preprocessing, we obtain a formal context suitable for the analysis of FCA algorithms. Details on the size of the formal context, the number of concepts, and the density of these contexts can be found in Table \ref{tab5}. The density represents the proportion of 1's in the formal context, calculated as the ratio of the number of 1's to the total number of elements (both 0's and 1's). The formula is expressed as:
\[
\text{Density} = \frac{n_1}{n_0 + n_1}
\]
where $ n_1 $ represents the number of 1's and $ n_0$ represents the number of 0's. The datasets, including Seeds, Mushroom, Nursery, Appliances Energy Prediction, Letter, Adult, Relation Network, and Health Indicators, are referred to as Dataset 1 through Dataset 7 for convenience.

\begin{table*}[ht]
	\centering
	\caption{ The processed datasets.}\label{tab5}
	\begin{tabular*}{\textwidth}{@{\extracolsep{\fill}}lccc@{\extracolsep{\fill}}}
		\toprule
		Datasets   & Size of formal contexts & Number of concepts & Density (\%) \\
		\midrule
		Dataset 1 & $8,124 \times 18$            &208         & 23.37          \\
		Dataset 2 & $12,960 \times 12$           &630         & 27.12          \\
		Dataset 3 & $19,735 \times 15$           &4,485       & 46.77          \\
		Dataset 4 & $20,000 \times 22$           &1,373       & 6.02           \\
		Dataset 5 & $48,842 \times 20$           &1,270       & 9.63           \\
		Dataset 6 & $53,413 \times 14$           &4,410       & 36.02          \\
		Dataset 7 & $253,680 \times 21$          &5,349       & 22.38 		   \\
		\bottomrule
	\end{tabular*}
\end{table*}

\subsection{Comparison of result amd solution performance}
By applying the concept construction algorithm and the homomorphic encryption-based concept construction (HECC) method to the datasets mentioned above, we observe from Figure 5 that the number of concepts constructed by the proposed HECC algorithm precisely matches the actual number of concepts. This further validates the accuracy and effectiveness of our approach.

\begin{figure}[H]
	\centering
	\includegraphics[width=0.60\textwidth]{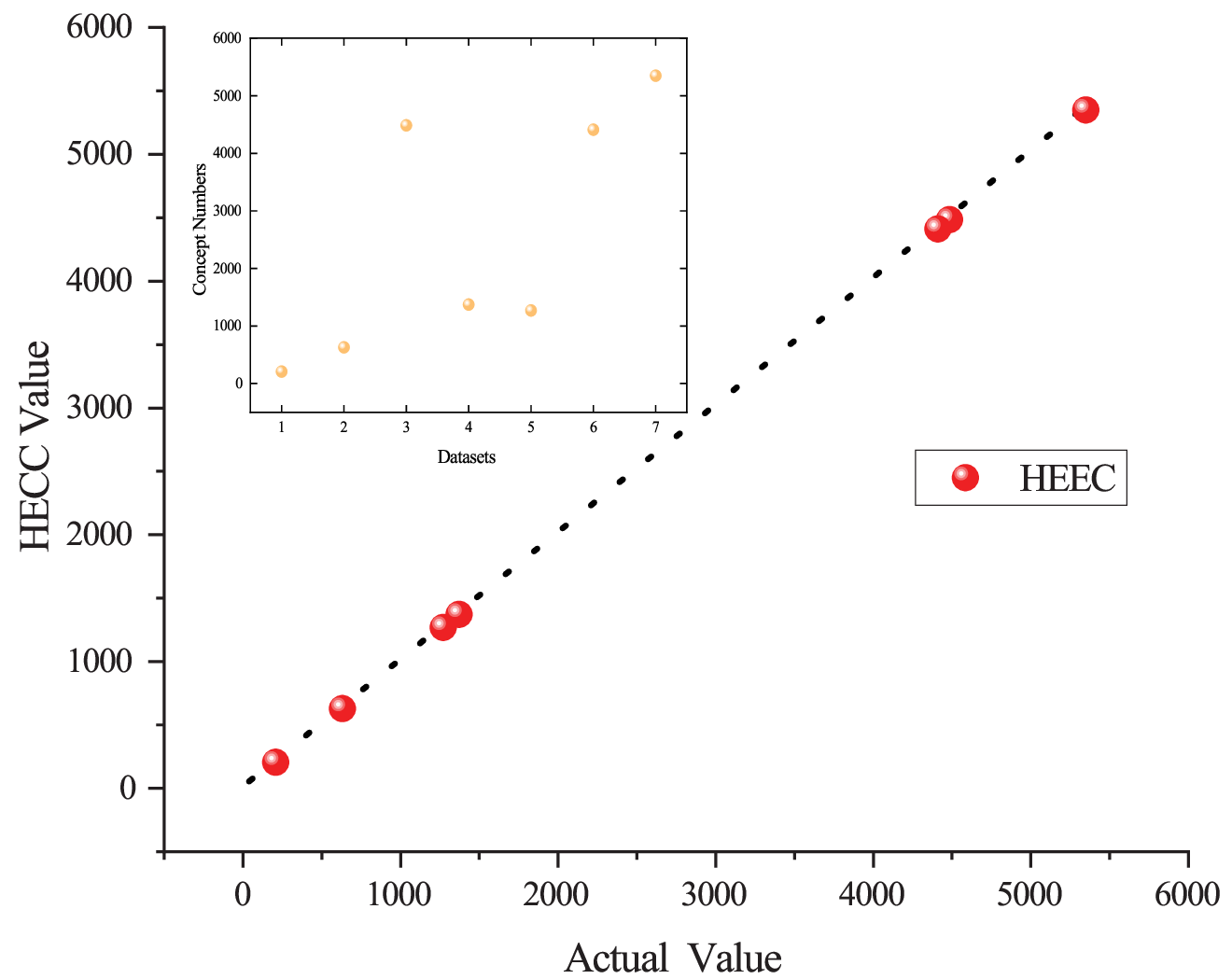}
	\caption{Comparison of HECC and the actual number of concepts.}
	\label{fig6}
\end{figure}

To analyze the efficiency of HECC, we compare the method proposed in this paper with the basic FCA concept construction methods, including the enumeration method (TEM) \cite{1} and the incremental algorithm (TIA) for constructing formal concepts \cite{34}.

From Table \ref{tab6}, we can observe that the HECC method constructs concepts in less time compared to the TEM and TIA methods. As the size of the dataset increases, the efficiency of the HECC method improves significantly. The time complexity of HECC method in this paper is greatly affected by the data structure, especially the number of attributes. For example, in Datasets 4 and Dataset 6, although the number of objects in Dataset 6 is much larger than in Dataset 4, the number of attributes in Dataset 6 is smaller. However, the running time of Dataset 6 is significantly higher than that of Dataset 4.

\begin{table*}[ht]
	\centering
	\caption{Comparison of the generation time for formal concepts of different algorithms (Unit: s).}\label{tab6}
	\begin{tabular*}{\textwidth}{@{\extracolsep{\fill}}lccc@{\extracolsep{\fill}}}
		\toprule
		Dataset   & HECC           & TEM     & TIA         \\
		\midrule
		Dataset 1 & 41.07                    & 217.34       & 46.00         \\
		Dataset 2 & 4.84                     & 40.82       & 356.00 		\\
		Dataset 3 & 34.98                    & 53.58       & 6,486.00	    \\
		Dataset 4 & 1,715.39                 & 9,525.67       & 2,216.00 		    \\
		Dataset 5 & 1,227.63                 & 3,676.85       & 4,495.00 		    \\
		Dataset 6 & 44.54                    & 77.08       & 45,658.00 	    	\\
		Dataset 7 & 14,925.14                & 75,869.60       & 1,296,000.00 	     	\\
		\bottomrule
	\end{tabular*}
\end{table*}

As demonstrated in Tables \ref{tab7} and Tables \ref{tab8}, the HECC exhibits superior performance as the number of objects or attributes increases. Table \ref{tab9} indicates that, when the density of the formal context is relatively low, the HECC and EM outperforms TIA and remains the optimal choice even in high-density contexts. This can be attributed to the fact that the HECC is a fundamental traversal-based approach, which is not significantly affected by increases in density.

{	
	\vspace{1em} 
	
	\noindent
	\captionof{table}{The running time with the number of objects($|U|$) increasing from 10000 to 50000 with an interval of 10000, where $|A|= 15$ and $p=0.10$.}\label{tab7}
	\begin{tabular*}{\linewidth}{@{\extracolsep{\fill}}c c c c c c}
		\toprule
		\multirow{2}{*}{Method} & \multicolumn{5}{c}{Object} \\
		\cmidrule{2-6}
		& 10,000 & 20,000 & 30,000 & 40,000 & 50,000 \\
		\midrule
		HECC  &8.64 & 20.24 & 31.17 & 42.35  &55.87 \\
		EM & 24.00  & 49.21  & 73.77  & 97.22 & 122.88 \\
		TIA & 219.00 & 1,104.00 & 3,028.00 &6,161.00 & 10,825.00 \\
		\bottomrule
	\end{tabular*}

	\vspace{1em} 
	
	\noindent
	\captionof{table}{The running time with the number of objects($|A|$) increasing from 14 to 22 with an interval of 2, where $|U|= 50000$ and $p=0.10$.}\label{tab8}
	\begin{tabular*}{\linewidth}{@{\extracolsep{\fill}}c c c c c c}
		\toprule
		\multirow{2}{*}{Method} & \multicolumn{5}{c}{Attribute} \\
		\cmidrule{2-6}
		& 14 & 16 & 18 & 20 & 22 \\
		\midrule
		HECC  & 55.03 & 200.04 & 536.23 & 955.45 & 1,214.63  \\
		EM & 57.60  & 258.34  & 1147.96  & 5006.16 & 21932.20 \\
		TIA & 10,825.00  & 12,030.00  & 11,954.00  & 11,612.00  & 12,297.00  \\
		\bottomrule
	\end{tabular*}
	
	\vspace{1em} 
	
	\noindent
	\captionof{table}{The running time with the density ($p$) increasing from 6$\%$ to 10$\%$ with an interval of 1$\%$, where $|U|= 50000$ and $|A|=15$.}\label{tab9}
	\begin{tabular*}{\linewidth}{@{\extracolsep{\fill}}c c c c c c}
		\toprule
		\multirow{2}{*}{Method} & \multicolumn{5}{c}{Density} \\
		\cmidrule{2-6}
		& 6 $\%$ & 7$\%$ & 8$\%$ & 9$\%$ & 10$\%$ \\
		\midrule
		HECC  & 54.23 & 54.36 & 54.38 & 54.45 & 54.76 \\
		EM & 123.54  & 125.24  & 123.02  & 122.96 & 125.41 \\
		TIA & 8,545.00  & 9,514.00  & 9,635.00  & 10,078.00  & 10,825.00  \\
		\bottomrule
	\end{tabular*}
}
	
The main steps of the experiments in this paper include matrix reading, encryption, calculating the attribute power set, and constructing privacy-preserving concept. We record the time taken for each part. As shown in Table \ref{tab10}, we find that calculating the attribute power set takes the longest time.

\begin{table*}[ht]
	\centering
	\caption{Analysis of the running time of each stage of the HECC (Unit: s).}\label{tab10}
	\begin{tabular*}{\textwidth}{@{\extracolsep{\fill}}lccccc@{\extracolsep{\fill}}}
		\toprule
		Dataset   & Total     & Matrix reading     & Encryption      & Processing  & privacy-preserving concept    \\
		\midrule
		Dataset 1 & 41.07           & 0.01      & 3.02                & 38.25 & 0.006  \\
		Dataset 2 & 4.84            & 0.01      & 3.21		          & 1.15  & 0.003 \\
		Dataset 3 & 34.98         & 0.02         & 6.11	              &  22.90 & 5.48  \\
		Dataset 4 & 1,715.39        & 0.03       & 9.08		          & 1,705.80   & 0.019   \\
		Dataset 5 & 1,227.63       & 0.03     & 20.32		          & 1,206.80    & 0.017 \\
		Dataset 6 & 44.54          & 0.05        & 15.44 	          & 28.22  & 0.243 	\\
		Dataset 7 & 14,925.14     & 0.21             & 109.89 	      & 14,813.85   & 0.223 	\\
		\bottomrule
	\end{tabular*}
\end{table*}

\subsection{The parallel implementation of HECC}
From the above analysis, it can be seen that the algorithm's running time is primarily concentrated in the power set calculation part. To improve the algorithm's efficiency, we propose a parallel version of the HECC algorithm, inspired by the principles of parallel algorithms. The following section will provide a detailed description of the parallelization method.

As shown in Figure 6, the attribute power set candidate subsets are constructed through thread-level parallel computation during the subset generation phase. In the computation-intensive support calculation phase, a dynamic task assignment strategy is employed, combined with thread-local hash table storage and invalid subset pruning techniques, significantly improving computational efficiency.

\begin{figure}[H] 
	\centering
	\includegraphics[width=0.60\textwidth]{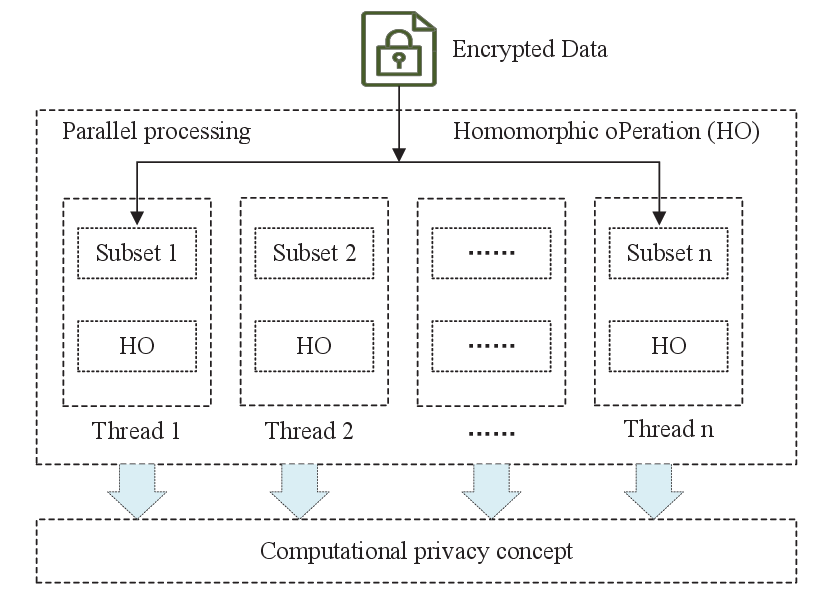}
	\caption{The relative running time of parallel HECC.}
	\label{fig6}
\end{figure}

Figure 7 illustrates that as the number of threads increases, the relative runtime of most datasets decreases significantly, demonstrating the effectiveness of parallel computation in reducing execution time. However, the performance under parallel computation is influenced by the characteristics of each dataset. For example, the Mushroom dataset shows a sharp decline in runtime as the number of threads increases, reflecting a substantial performance improvement. In contrast, the Nursery and Health Indicators datasets exhibit a more gradual decrease in runtime, with performance improvements remaining relatively stable. The Appliances Energy Prediction dataset shows a mild decline, suggesting a consistent parallelization effect. In comparison, the Letter and Adult datasets experience slower reductions in runtime, indicating that the performance gains from parallel computation are more modest for these datasets.

\begin{figure}[H]
	\centering
	\includegraphics[width=0.60\textwidth]{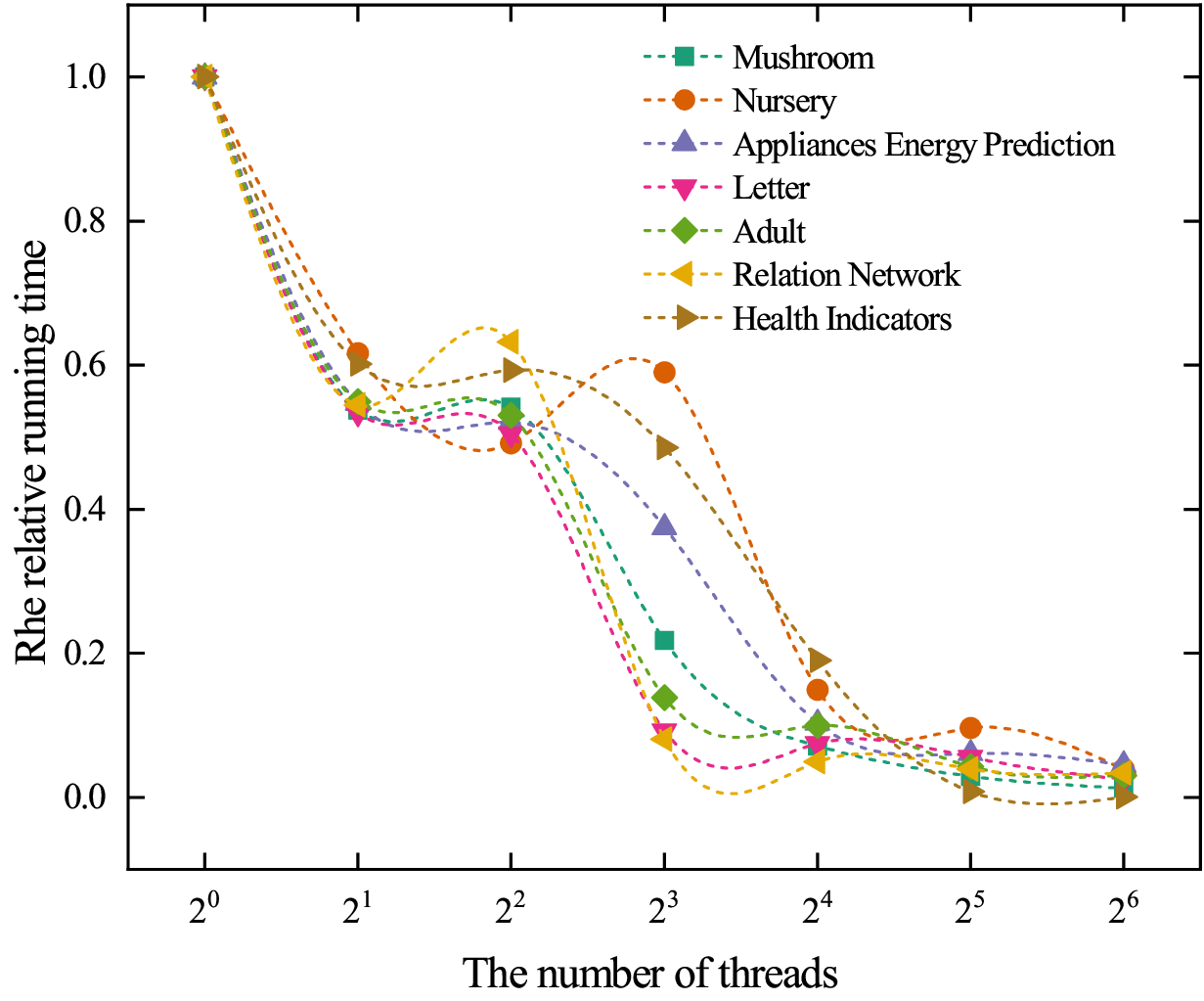}
	\caption{The relative running time of parallel HECC.}
	\label{fig6}
\end{figure}

\section{Conclusion and disscusion}
In this study, inspired by privacy-preserving computing techniques, we propose a homomorphic encryption-based method for the construction of privacy-preserving concepts. This approach enables users to safely construct concepts with the assistance of third-party platforms such as cloud computing, without disclosing their private information. In addition, we validate the security, efficiency, and feasibility of our method through experiments.

There are still some issues worth further research. The algorithm proposed in this study experiences significant increases in computation time when dealing with large attribute sets with high time complexity. Future work should focus on developing more efficient algorithms to address this limitation. An interesting research direction would be to explore methods for constructing models that enable knowledge discovery and data mining directly from privacy-preserving concepts.

\section*{Credit authorship contribution statement}

\textbf{Qiangqiang Chen}: Writing-original draft, Validation, Methodology. \textbf{Yunfeng Ke}: Software, Data curation. \textbf{Jinhai Li}: Writing-reviewing and editing, Funding acquisition. \textbf{Shen Li}: Investigation, Writing-reviewing and editing, Supervision.

\section*{Declaration of Competing Interest}

The authors declare that they have no known competing financial interests or personal relationships that could have appeared to influence the work reported in this paper.

\section*{Data availability}

Data will be made available on request (https://archive.ics.uci.edu/).

\section*{Acknowledgements}
This work was supported by the China National Natural Science Foundation under Grant 62476114, the Yunnan Fundamental Research Projects under Grant 202401AV070009 and the Yunnan Province Natural Science Foundation under Grant 202201AU070136.

\end{document}